\documentclass[twocolumn,10pt]{article}
\usepackage[utf8]{inputenc}
\usepackage[dvips]{graphicx}
\usepackage{amsthm}
\usepackage{amsmath}
\usepackage{amssymb}
\usepackage{natbib}
\usepackage{enumerate}
\usepackage[dvips]{graphicx}
\usepackage[usenames,dvipsnames]{xcolor}

\newcommand{\ud}{\mathrm{d}}
\newcommand{\EdStime}{t_0^{\scriptscriptstyle{(EdS)}}}

\title{Construction of the cosmological model with periodically distributed inhomogeneities with growing amplitude.}
\author{Szymon Sikora, Krzysztof Głód \\
\scriptsize{\textit{Astronomical Observatory, Jagiellonian University, Orla 171, 30-244 Kraków, Poland}} }
  
\begin{document}

\twocolumn[
\begin{@twocolumnfalse}
\maketitle
\begin{abstract}
We construct an approximate solution to the cosmological perturbation theory around Einstein-de Sitter background up to the fourth-order perturbations. This could be done with the help of the specific symmetry condition imposed on the metric, from which follows, that the model density forms an infinite, cubic lattice. We show that the perturbative solution obtained this way can be interpreted as the exact solution to the Einstein equations for a dust-like energy-momentum tensor. In our model, it seems that physical quantities averaged over a large scales overlap with the respective Einstein-de Sitter prediction, while local observables could differ significantly from their background counterparts. As an example, we analyze in details a behaviour of the local and the global measurements of the Hubble constant, which is important in the context of a current Hubble tension problem.
\end{abstract}
\vspace{1cm}
\end{@twocolumnfalse}
]

\small
\section{Introduction}
The studies within the cosmological perturbation theory up to second order yield that the influence of inhomogeneities on the Hubble diagram is visible but small and can reach at most one percent level in the parameters estimation \cite{2013PhRvL.110b1301B,2013JCAP...06..002B,2014PhRvL.112v1301B,2014JCAP...11..036C,2015JCAP...06..050B,2015JCAP...07..040B,2017JCAP...03..062F}. Similar results are reached by ray tracing into Newtonian N-body numerical simulations \cite{2013PhRvD..88j3527A,2014CQGra..31w4006A,2015PhRvL.114e1302A}, relativistic N-body numerical simulations \cite{2016NatPh..12..346A,2016JCAP...07..053A,2018PhRvD..97d3509E,2019PhRvD.100b1301A} or relativistic hydrodynamical numerical simulations of an inhomogeneous dust model \cite{2016PhRvL.116y1301G,2016PhRvD..93l4059M,2016ApJ...833..247G,2017PhRvD..95f4028M,2018ApJ...865L...4M,2019PhRvD..99f3522M}. It is suggested that the emergence of spatial curvature during structure formation could be at least partially responsible for this effect \cite{2017JCAP...02..047S,2018CQGra..35b4003B,2018PhRvD..97j3529B,2019JCAP...05..039C,2019arXiv190504588C}.

The negligible impact of inhomogeneities on the global evolution and observational parameters of cosmological models is supported by general considerations concerning the backreaction effect \cite{2014CQGra..31w4003G,2015arXiv150606452G,2016CQGra..33l5027G,2017PhRvD..95l4009F,2017MNRAS.469..744K,2019CQGra..36a4001A} and perturbative analysis of weak gravitational lensing \cite{2016MNRAS.455.4518K}. However, these approaches are often criticized as incomplete or inconclusive because of restrictive assumptions made \cite{2015CQGra..32u5021B,2018MNRAS.473L..46B,2018arXiv180609530E,2018CQGra..35xLT02B}. There are also works which do not confirm the irrelevance of inhomogeneities despite similar methods and techniques used in studies \cite{2016PhRvL.116y1302B,2017MNRAS.469L...1R,2020JCAP...02..017F}. Moreover, it is argued that the phenomena of virialization of clusters and volume dominance of voids significantly affect the Hubble diagram and may even explain dark energy \cite{2013JCAP...10..043R,2018A&A...610A..51R,2019IJMPD..2850143D,2019CQGra..36q5006V,2020arXiv200210831H}.

Investigations to light propagation in inhomogeneous cosmologies have brought the development of various constructions for the models. They include a~Swiss-cheese arrangement of the Lema\^itre-Tolman or the Szekeres holes into the Robertson-Walker background space-time \cite{2013PhRvD..87l3526F,2013JCAP...12..051L,2014PhRvD..90l3536P,2015JCAP...10..057L,2017PhRvD..95f3532K,2019PhRvD.100f3533K} or a~regular lattice of the Schwarzschild spheres approximately glued by the Lindquist-Wheeler technique \cite{2009JCAP...10..026C,2009PhRvD..80j3503C,2011PhRvD..84j9902C,2012PhRvD..85b3502C,2012PhRvD..86d4027Y,2013PhRvL.111p1102Y,2014PhRvD..89l3502Y,2015PhRvD..92f3529L}. With the help of numerical simulations, it is also possible to study evolving configurations consisting of interacting black holes \cite{2012CQGra..29p5007B,2012PhRvD..86d3506C,2013JCAP...11..010C,2013CQGra..30w5008B,2014CQGra..31h5002K,2017CQGra..34f5009D,2017JCAP...03..014B,2017JCAP...10..012D,2018CQGra..35q5004B,2019CQGra..36s5009G}. An exceptional example of model with point masses periodically distributed on a~cubic lattice, presented in \cite{2012CQGra..29o5001B,2013CQGra..30b5002B}, can be built as a~perturbative solution to the Einstein equations for a~dust. The post-Newtonian formalism is another framework utilizing a~perturbative approach in which a~model is hierarchically constructed from autonomous cells whose matching conditions determine an overall dynamics \cite{2015PhRvD..91j3532S,2016PhRvD..93h9903S,2016PhRvD..94b3505S,2017CQGra..34f5003S,2017JCAP...07..028S,2020PhRvD.101f3530C}.

Recently, we have proposed a~cosmological model containing a~continuous periodic distribution of inhomogeneities which are characterized by the equation of state of a~dust \cite{2017PhRvD..95f3517S,2019PhRvD..99h3521S}. The space-time of the model is constructed perturbatively and the matter is treated within the fluid hydrodynamic formalism. Here, we present a~substantial extension of this model.

\section{Model construction}\label{sec:construction}
Consider the following space-time metric in the Cartesian-like coordinates $(t,x,y,z)$:
\begin{equation}\label{eqn:full_metric}
g_{\mu\:\!\nu}=\begin{pmatrix}
-1 & 0 & 0 & 0 \\
0 & c_{11} & c_{12} & c_{13} \\
0 & c_{12} & c_{22} & c_{23} \\
0 & c_{13} & c_{23} & c_{33}
\end{pmatrix}\,,
\end{equation}
where the six metric functions $c_{ij}=c_{ij}(\lambda,t,x,y,z)$ depend on the space-time coordinates and some auxiliary parameter $\lambda$. We adopt geometrized units for which $c=1$ and $G=1$. Since the Christoffel symbols $\Gamma^\mu_{0\:\!0}=0$ vanish for all $\mu$, the vector $U^\mu=(1,0,0,0)$ is always tangent to some time-like geodesic. The worldlines of dust particles are geodesics, therefore for the universe filled with dust, the coordinates we use are comoving.

The task we want to address in this article is the following. Suppose that the Einstein equations are satisfied exactly. How to find the metric functions $c_{ij}$ so that the energy-momentum tensor $T_{\mu\:\!\nu}=G_{\mu\:\!\nu}/(8\pi)$ is a dust-like energy-momentum tensor and the density distribution has inhomogeneities with the amplitude growing during the time evolution? 

By the \emph{dust-like} energy-momentum tensor we mean such $T_{\mu\:\!\nu}$ for which all of the elements, except the density $\rho=T_{0\:\!0}$, are negligible compared to $\rho$. A variety of exact solutions to the Einstein equations, which tend to describe the Universe in the matter-dominated era, assume the dust energy-momentum tensor for simplicity. In that case, all of the $T_{\mu\:\!\nu}$ elements other than the density are exactly zero. However, many physical processes could contribute to these energy-momentum tensor elements. For example, proper motions of the galaxies act as the pressure, while this contribution is very small. This justifies the dust-like assumption for $T_{\mu\:\!\nu}$ in the matter-dominated era if only smallness of the elements other than the density is properly controlled.

The task of finding metric functions $c_{ij}$ for which the energy-momentum tensor has the properties described above is quite complicated in general. To handle it in the specific case we consider two simplifying assumptions:
\begin{enumerate}[(i)]
\item The metric functions $c_{ij}$ are invariant over every permutation of the spatial variables $(x,y,z)$.
\item The metric functions $c_{ij}$ are analytic in $\lambda$ and the parameter $\lambda$ is proportional to the amplitude of the inhomogeneities. Additionally, we assume that for $\lambda=0$ we recover the Einstein-de Sitter (EdS) space-time, which is the spatially flat universe homogeneously filled with dust. This means that the metric functions in this limit reads: $c_{ij}(0,t,x,y,z)=a(t)^2$ for $i=j$ and zero otherwise. The scale factor in this case is $a(t)=\mathcal{C}\,t^{2/3}$, where $\mathcal{C}$ is a constant.
\end{enumerate}

The assumption (ii) enables us to consider the usual perturbation theory around the EdS background.
We may take Taylor expansion of the metric and resulting energy-momentum tensor around $\lambda=0$:
\begin{equation}
 g_{i\:\!j}=\sum\limits_{k=0}^{\infty} \lambda^k\,g^{(k)}_{i\:\!j}\,, \quad T_{\mu\:\!\nu}=\sum\limits_{k=0}^{\infty} \lambda^k\,T^{(k)}_{\mu\:\!\nu}\,,
\end{equation}
and analyze $k$-order metric elements $g^{(k)}_{i\:\!j}$ and $k$-order energy-momentum tensor $T^{(k)}_{\mu\:\!\nu}$ order by order. The form of the metric (\ref{eqn:full_metric}) implies that perturbations are performed in the synchronous comoving gauge in each order. The scalar perturbations in the linear order admit two modes: the decaying and the growing one. In our prevuous paper \cite{2019PhRvD..99h3521S} we analyze the decaying mode consistent with the assumption (i) in the orders higher than the linear order. In the current paper we concentrate on the growing mode. We will start with the linear perturbations and then we correct the solution in the consecutive orders. At the end of this procedure, we get the desired exact dust-like solution, finding of which is the main goal of the present paper.

\subsection{Perturbation theory in the linear order.}\label{sec:lin}
We assume the following ansatz for the spatial part of the metric in the $k$-order, which is consistent with the symmetry assumption (i):
\begin{multline}\label{eqn:metric_form}
g^{(k)}_{i\:\!j}=t^{\alpha_k}\,a(t)^2
\begin{pmatrix}
A_x^{(k)} &0&0\\
0&A_y^{(k)}&0\\
0&0&A_z^{(k)}
\end{pmatrix}+ \dots \\
\dots+ t^{\beta_k}\,a(t)^2
\begin{pmatrix}
B_{x\:\!y\:\!z}^{(k)} &0&0\\
0&B_{x\:\!y\:\!z}^{(k)}&0\\
0&0&B_{x\:\!y\:\!z}^{(k)}
\end{pmatrix}+ \dots  \\
\dots+ t^{\phi_k}\,a(t)^2
\begin{pmatrix}
0&F_{x\:\!y}^{(k)}&F_{z\:\!x}^{(k)}\\
F_{x\:\!y}^{(k)}&0&F_{y\:\!z}^{(k)}\\
F_{z\:\!x}^{(k)}&F_{y\:\!z}^{(k)}&0
\end{pmatrix}\,,
\end{multline}
where the following abbreviated notation is used: $A_i^{(k)}\equiv A^{(k)}(x^i)$, $B_{x\:\!y\:\!z}^{(k)}\equiv B^{(k)}(x)+B^{(k)}(y)+B^{(k)}(z)$ and $F_{i\:\!j}^{(k)} \equiv F^{(k)}(x^i,x^j)$. In each order we introduced three functions $A^{(k)}$, $B^{(k)}$, $F^{(k)}$ and three coefficients $\alpha_k$, $\beta_k$ and $\phi_k$, which should be specified in the process of the model construction. The symmetry condition (i) implicates that the energy-momentum tensor in each order $k\geq 1$ has the four types of components: $T^{(k)\:\!0}\,{}_0$, $T^{(k)\:\!0}\,{}_i$, $T^{(k)\:\!i}\,{}_j$ for $i=j$ and $T^{(k)\:\!i}\,{}_j$ for $i\neq j$. The structural form of the components within each type is invariant over every permutation of spatial variables. 

Let us start with the linear order perturbations. If we specify the powers:
\begin{equation}
\alpha_1=2/3\,, \quad \beta_1=0\,,\quad \phi_1=0\,,
\end{equation} 
then the elements $T^{(1)\:\!0}\,{}_i$ and $T^{(1)\:\!i}\,{}_j|_{i\neq j}$ are equal to zero and all of the terms of $T^{(1)\:\!0}\,{}_0$ together with all of the terms of $T^{(1)\:\!i}\,{}_j|_{i=j}$ has a simple power law dependence on time. For simplicity, we put the function $F^{(1)}=0$. Then, we may cancel out the pressure-like terms \mbox{$T^{(1)\:\!i}\,{}_j|_{i=j}=0$} demanding that the function $B^{(1)}$ satisfies differential equation:
\begin{equation}\label{eqn:B1}
\frac{\ud^2}{\ud w^2}B^{(1)}(w)=\frac{10}{9}\,\mathcal{C}^2\,A^{(1)}(w)\,.
\end{equation}
After that, the first order density $\rho^{(1)}\equiv -T^{(1)\:\!0}\,{}_0$ is:
\begin{equation}\label{Eqn:rho1}
\rho^{(1)}=\frac{-1}{12\pi t^{4/3}}\,\left(A^{(1)}(x)+A^{(1)}(y)+A^{(1)}(z) \right)\,.
\end{equation}
This way we obtain exact dust solution to the cosmological perturbation theory in the linear order, in which the density distribution in space is given by the arbitrary function $A^{(1)}$. Let us analyze in details the model for one exemplary function $A^{(1)}$ given below.

In the beginning, we comment on the dimensions of the physical quantities important for model construction. 
We are working in the geometrized units $c=1$ and $G=1$. In this system of units, all of the physical quantities are dimensionless or their dimension can be expressed as some power of the unit of length. 
We choose the megaparsec as basic unit of length. Then, the age of the EdS universe $\EdStime=9.32\,\mathrm{Gyr}$, compatibile with the Hubble constant value $H_0=70\,\mathrm{km/s/Mpc}$, reads $\EdStime=2855.57\,\mathrm{Mpc}$. There is a natural convention of normalizing the scale factor to unity at the age of the universe $a(\EdStime)=1$. From that follows the value of the constant $\mathcal{C}=4.97\times 10^{-3}$. The density of the EdS model $\rho^{(0)}=1/(6\pi t^2)$ evaluated at the universe age defines the critical density $\rho_{cr}=\rho^{(0)}(\EdStime)$, which value is $\rho_{cr}=6.51\times 10^{-9}\,\mathrm{Mpc}^{-2}$. The critical density introduces a natural density scale. We define the quantity $\Omega:=\rho/\rho_{cr}$ as a density measured in the critical density units.

\begin{figure}[h]
	\centering
	\includegraphics[width=0.3\textwidth]{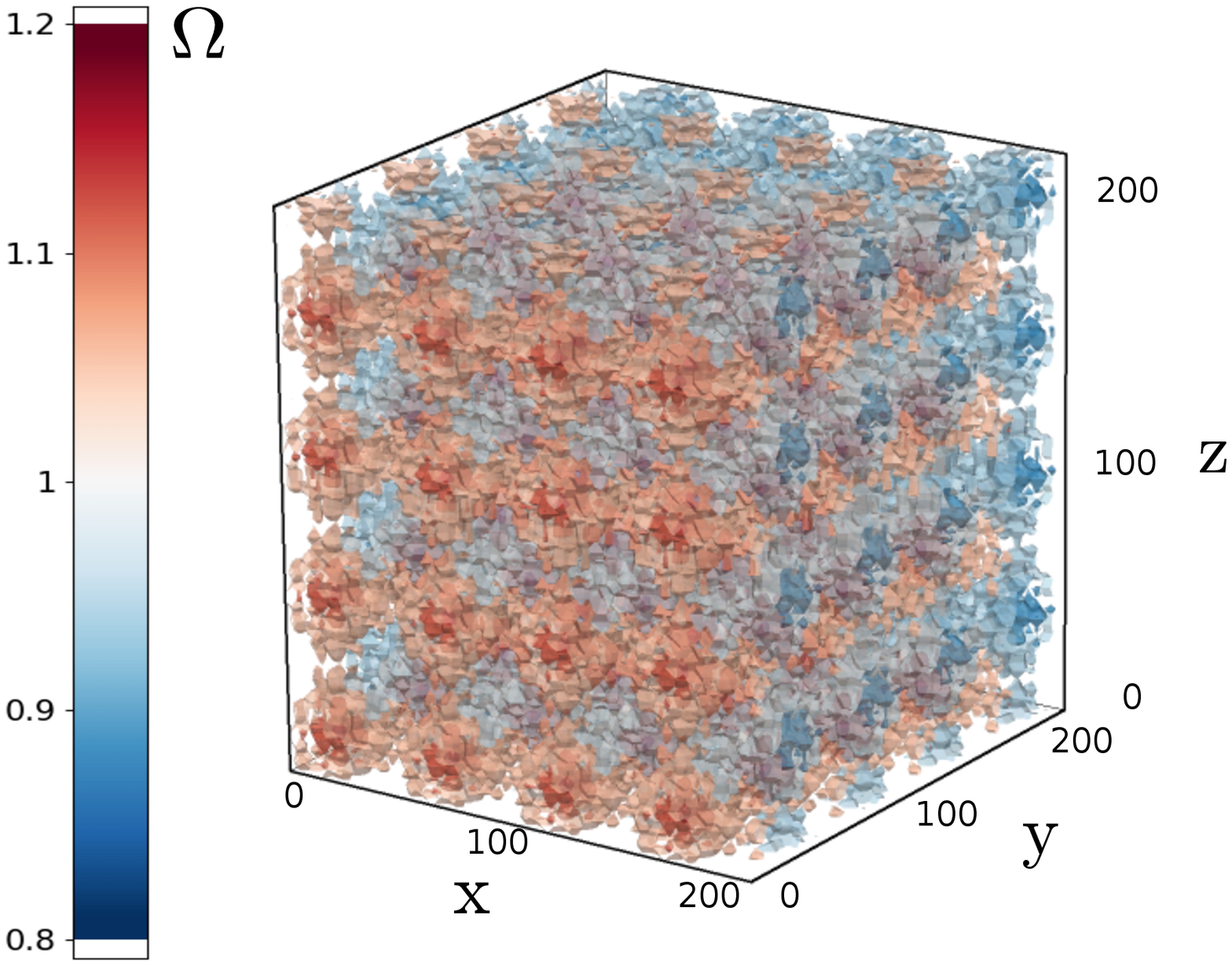}
   \includegraphics[width=0.3\textwidth]{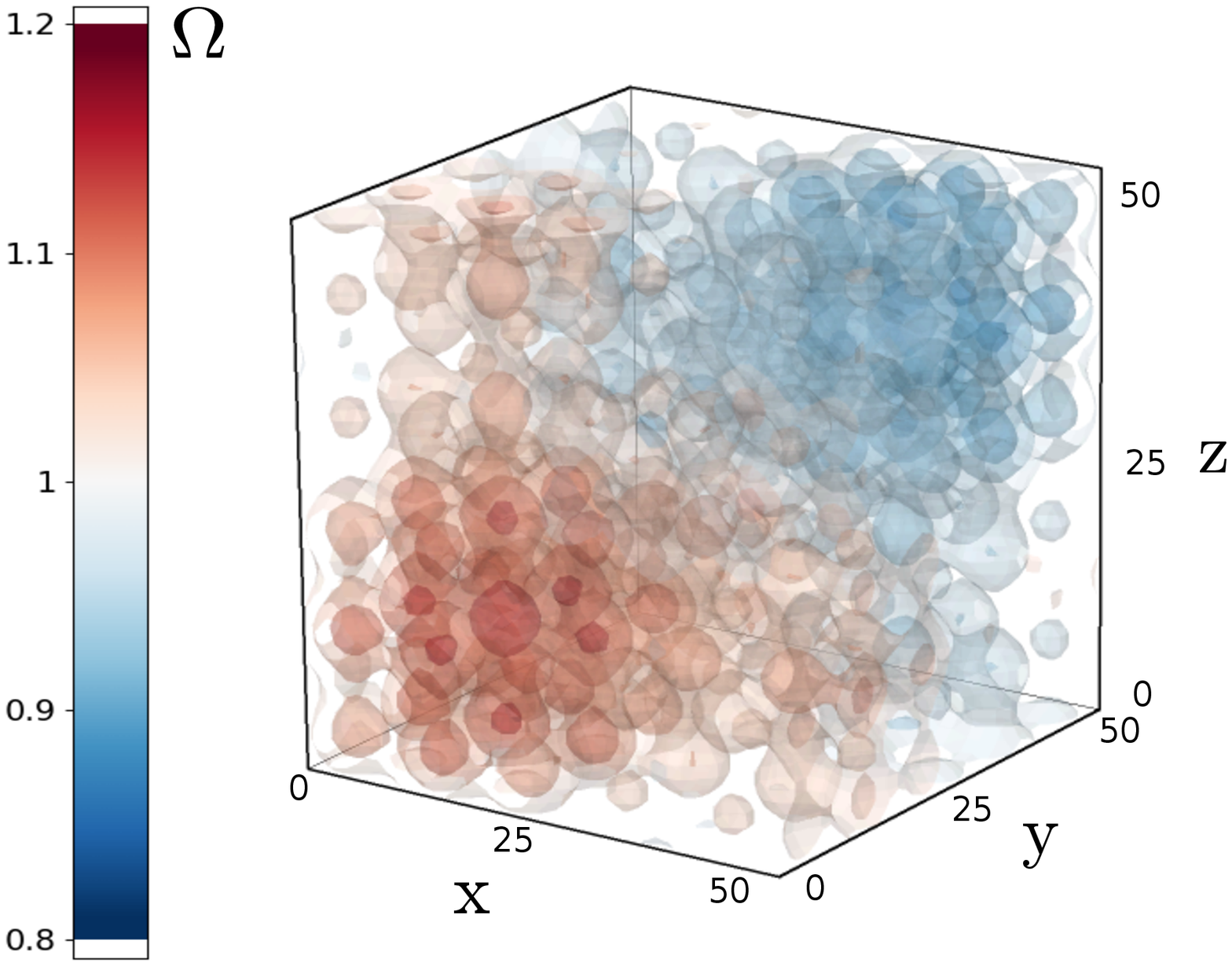}
	\caption{\label{fig:isodensity} \scriptsize{Top: The model isodensity surfaces at $\EdStime$, which form an infinite, periodic lattice. Bottom: The density distribution within the elementary cell.}}
\end{figure}

Now, we choose the function $A^{(1)}$ as:
\begin{equation}\label{Eqn:A1}
A^{(1)}(w)=-s_0\,\sin(\mathcal{B}_0\,w)-s_1\,\sin(\mathcal{B}_1\,w)\,,
\end{equation}
where $s_0=1$, $s_1=0.5$, $\mathcal{B}_0=\pi/25$ and $\mathcal{B}_1=\pi/5$. If we fix the lambda parameter value 
as $\lambda=4.42\times 10^{-4}$, then the maximal density at $\EdStime$ is $\Omega=1.2$. Density distribution illustrated in 
\mbox{Fig. \ref{fig:isodensity}} forms periodic, cubic lattice.
The linear size of the elementary cell at $\EdStime$ is around $50\,\mathrm{Mpc}$. Because the metric is nontrivial, the length of a segment
depends on its orientation and position in space. The linear size of the elementary cell measured near to the overdense region
gives the value slightly lower than $50\,\mathrm{Mpc}$, while the result of the same measurement performed close to the underdense region 
could be slightly higher than $50\,\mathrm{Mpc}$. For a scales much larger than the size of the elementary cell the model universe becomes homogeneous and isotropic in common sense, and FLRW space-time arises as a natural candidate for an effective average model. Although the density distribution profile is restricted by the model symmetry, quite complicated distributions are allowed. In the given example, one can identify large overdensity and underdensity regions of a size comparable to the typical size of superclusters of galaxies and smaller substructures with a size around a few megaparsecs. 

\subsection{Perturbation theory in the second order.}
The function $B^{(1)}$ which is a solution of the Eqn. \ref{eqn:B1} for the above proposition of the arbitray function $A^{(1)}$ is:
\begin{equation}
B^{(1)}(w)=\frac{10\,\mathcal{C}^2}{9}\left(\frac{s_0}{\mathcal{B}_0^2}\sin(\mathcal{B}_0\,w)+\frac{s_1}{\mathcal{B}_1^2}\sin(\mathcal{B}_1\,w) \right)\,.
\end{equation}
In the above formula there appears two small constants: $(\mathcal{C}/\mathcal{B}_0)^2=1.56\times 10^{-3}$ and $(\mathcal{C}/\mathcal{B}_1)^2=6.25\times 10^{-5}$, where $\mathcal{C}^2=2.47\times{10}^{-5}$. The metric function $B^{(1)}$ is then much smaller than the metric function $A^{(1)}$. However, in the second-order energy-momentum tensor elements appear some terms containing derivatives of the function $B^{(1)}$ divided by $\mathcal{C}^2$. These terms are the same order of magnitude as the terms with the function $A^{(1)}$ alone. For that reason, one cannot neglect the metric function $B^{(1)}$ in the first-order metric $g^{(1)}$, if the second-order energy-momentum tensor is considered. Nevertheless, the existence of these small constants enables one to identify the leading terms in the second-order energy-momentum tensor and to neglect the terms which are much smaller in comparison to the leading terms.
  
From now on, we denote by the symbol $\approx$ the approximation of some expression, in which all of the terms proportional to $\mathcal{C}^2$
are neglected. Because some subexpressions could have a different time dependence, the validity of this approximation should be checked in each epoch of time.  Lets write the leading terms of $T^{(2)\:\!x}\,{}_x$.
\begin{multline}\label{Eqn:T2xx}
T^{(2)\:\!x}\,{}_x \approx \frac{A^{(1)}\left(y\right)^{2}}{12 \, \pi t^{\frac{2}{3}}} - \frac{A^{(1)}\left(y\right) A^{(1)}\left(z\right)}{72 \, \pi t^{\frac{2}{3}}} + \frac{A^{(1)}\left(z\right)^{2}}{12 \, \pi t^{\frac{2}{3}}} - \dots  \\
\dots-\frac{7 \, A^{(2)}\left(y\right)}{36 \, \pi t^{\frac{2}{3}}}- \frac{7 \, A^{(2)}\left(z\right)}{36 \, \pi t^{\frac{2}{3}}} - \dots \\
\dots- \frac{\frac{\ud}{\ud y}A^{(1)}\left(y\right) \frac{\ud}{\ud y}B^{(1)}\left(y\right)}{32 \, \pi {\mathcal{C}}^{2} t^{\frac{2}{3}}} - \frac{A^{(1)}\left(y\right) \frac{\ud^{2}}{\ud y^{2}}B^{(1)}\left(y\right)}{16 \, \pi {\mathcal{C}}^{2} t^{\frac{2}{3}}} - \dots \\
\dots- \frac{A^{(1)}\left(z\right) \frac{\ud^{2}}{\ud y^{2}}B^{(1)}\left(y\right)}{16 \, \pi {\mathcal{C}}^{2} t^{\frac{2}{3}}} - \frac{\frac{\ud}{\ud z}A^{(1)}\left(z\right) \frac{\ud}{\ud z}B^{(1)}\left(z\right)}{32 \, \pi {\mathcal{C}}^{2} t^{\frac{2}{3}}} - \dots \\
\dots- \frac{A^{(1)}\left(y\right) \frac{\ud^{2}}{\ud z^{2}}B^{(1)}\left(z\right)}{16 \, \pi {\mathcal{C}}^{2} t^{\frac{2}{3}}} - \frac{A^{(1)}\left(z\right) \frac{\ud^{2}}{\ud z^{2}}B^{(1)}\left(z\right)}{16 \, \pi {\mathcal{C}}^{2} t^{\frac{2}{3}}} +  \dots \\
\dots +\frac{\frac{\ud^{2}}{\ud y^{2}}B^{(2)}\left(y\right)}{16 \, \pi {\mathcal{C}}^{2} t^{\frac{2}{3}}} + \frac{\frac{\ud^{2}}{\ud z^{2}}B^{(2)}\left(z\right)}{16 \, \pi {\mathcal{C}}^{2} t^{\frac{2}{3}}} - \frac{\frac{\partial^{2}}{\partial y\partial z}F^{(2)}\left(y, z\right)}{8 \, \pi {\mathcal{C}}^{2} t^{\frac{2}{3}}}\,.
\end{multline}
In the beginning, time dependence of different terms was not the same. However, it 
can be simplified to a single power-law $t^{-2/3}$, when we fix the following values of the powers:
\begin{equation}
\alpha_2=4/3, \qquad \beta_2=2/3, \qquad \phi_2=2/3\,.
\end{equation}
If we wish that the terms containing the second-order metric functions $B^{(2)}$ and $F^{(2)}$ are the same order 
of magnitude as the terms containing functions $A^{(k)}$, then functions $B^{(2)}$ and $F^{(2)}$ should be 
proportional to $\mathcal{C}^2$. We will verify this assumption at the end of the presented procedure. 
In that case,  $T^{(2)\:\!0}\,{}_i\approx 0$ and $T^{(2)\:\!i}\,{}_j\approx 0$ for $i\neq j$. 
The corresponding elements $T^{(2)\:\!y}\,{}_y$ and $T^{(2)\:\!z}\,{}_z$ one would obtain by performing permutation of the spatial variables in the formula Eq. \ref{Eqn:T2xx}.

On the right-hand side of Eq. \ref{Eqn:T2xx} it is possible to separate terms depending on two variables. These terms are equal to zero when 
the following differential equation is satisfied:
\begin{multline}\label{Eqn:F2}
\frac{\partial^{2}}{\partial v\partial w}F^{(2)}\left(v, w\right) = -\frac{1}{9} \, {\mathcal{C}}^{2} A^{(1)}\left(v\right) A^{(1)}\left(w\right) - \dots \\
\dots - \frac{1}{2} \, A^{(1)}\left(w\right) \frac{\ud^{2}}{\ud v^{2}}B^{(1)}\left(v\right) - \frac{1}{2} \, A^{(1)}\left(v\right) \frac{\ud^{2}}{\ud w^{2}}B^{(1)}\left(w\right)\, .
\end{multline}
The remaining terms depending on one variable can be canceled out by the following condition:
\begin{multline}\label{Eqn:B2}
\frac{\ud^{2}}{\ud w^{2}}B^{(2)}\left(w\right) =  - \frac{4}{9}{\mathcal{C}}^{2} \, {\left(3 \, A^{(1)}\left(w\right)^{2} - 7 \, A^{(2)}\left(w\right)\right)}  + \dots \\
\dots+ \frac{1}{2} \, \frac{\ud}{\ud y}A^{(1)}\left(w\right) \frac{\ud}{\ud w}B^{(1)}\left(w\right) + A^{(1)}\left(w\right) \frac{\ud^{2}}{\ud w^{2}}B^{(1)}\left(w\right)\,.
\end{multline}
If the above differential equations are satisfied, then all of the elements $T^{(2)\:\!i}\,{}_j\approx 0$ 
for $i=j$. This is guaranteed by the symmetry condition imposed on the metric. 

The conditions Eq. \ref{Eqn:F2} and Eq. \ref{Eqn:B2} enable to simplify the form of the second-order density
 $\rho^{(2)}=-T^{(2)\:\!0}\,{}_0$. As the result, one gets:
\begin{multline}
\rho^{(2)}= \frac{A^{(1)}\left(x\right)^{2}}{9 \, \pi t^{\frac{2}{3}}} + \frac{A^{(1)}\left(y\right)^{2}}{9 \, \pi t^{\frac{2}{3}}} + \frac{A^{(1)}\left(z\right)^{2}}{9 \, \pi t^{\frac{2}{3}}} - \dots \\
\dots - \frac{5 \, A^{(2)}\left(x\right)}{18 \, \pi t^{\frac{2}{3}}} - \frac{5 \, A^{(2)}\left(y\right)}{18 \, \pi t^{\frac{2}{3}}} - \frac{5 \, A^{(2)}\left(z\right)}{18 \, \pi t^{\frac{2}{3}}}\,.
\end{multline}
Specification of the arbitrary function $A^{(2)}$ determines the second-order density. In our example, we choose
this function such that perturbed density at the second order has the same spatial distribution as the first-order 
density perturbation:
\begin{equation}
\rho^{(2)}=-\frac{\mathcal{K}}{t^{2/3}}\,\left(A^{(1)}(x)+A^{(1)}(y)+A^{(1)}(z) \right)\,.
\end{equation}
We introduced here the parameter $\mathcal{K}$, which will control the growth rate of the density contrast. In this
case, the function $A^{(2)}$ takes the form:
\begin{equation}
A^{(2)}(w)=\frac{18}{5}\,\pi\,\mathcal{K}\,A^{(1)}(w)+\frac{2}{5}\,A^{(1)}(w)^2\,.
\end{equation}

The right-hand sides of equations Eq. \ref{Eqn:F2} and Eq. \ref{Eqn:B2} depend only on the first-order metric
 functions and the arbitrary function $A^{(2)}$. In the case of our example, these functions are composed of the trigonometric 
functions and it is very simple to find solutions to Eq. \ref{Eqn:F2} and Eq. \ref{Eqn:B2}.
 Taking the constants of integration equal to zero for simplicity we obtained:
\begin{multline}
F^{(2)}\left(v, w\right) = -\frac{11\,\mathcal{C}^2}{9}\,\left( \frac{{s}_{0}^{2} \cos\left({\mathcal{B}}_{0} v\right) \cos\left({\mathcal{B}}_{0} w\right)}{  {\mathcal{B}}_{0}^{2}} + \right. \dots \\
\dots +\frac{ {s}_{0} {s}_{1} \cos\left({\mathcal{B}}_{1} v\right) \cos\left({\mathcal{B}}_{0} w\right)}{ {\mathcal{B}}_{0} {\mathcal{B}}_{1}} + \frac{ {s}_{0} {s}_{1} \cos\left({\mathcal{B}}_{0} v\right) \cos\left({\mathcal{B}}_{1} w\right)}{ {\mathcal{B}}_{0} {\mathcal{B}}_{1}} + \dots \\
\left. \dots+ \frac{ {s}_{1}^{2} \cos\left({\mathcal{B}}_{1} v\right) \cos\left({\mathcal{B}}_{1} w\right)}{ {\mathcal{B}}_{1}^{2}} \right)
\end{multline}
and
\begin{multline}
B^{(2)}\left(w\right) = \frac{7}{60} \, {\mathcal{C}}^{2} ({s}_{0}^{2}+{s}_{1}^{2}) w^{2} + \dots \\
\dots + \frac{{\mathcal{C}}^{2} {s}_{0} {s}_{1}\,\cos\left({\mathcal{B}}_{0} w + {\mathcal{B}}_{1} w\right)}{
 {\left({\mathcal{B}}_{0}^{2} + 2 \, {\mathcal{B}}_{0} {\mathcal{B}}_{1} + {\mathcal{B}}_{1}^{2}\right)}} \left(\frac{46}{45}+\frac{5\,\mathcal{B}_0}{18\,\mathcal{B}_1}+\frac{5\,\mathcal{B}_1}{18\,\mathcal{B}_0} \right)+\dots \\
\dots + \frac{{\mathcal{C}}^{2} {s}_{0} {s}_{1}\,\cos\left({\mathcal{B}}_{0} w - {\mathcal{B}}_{1} w\right)}{
 {\left({\mathcal{B}}_{0}^{2} - 2 \, {\mathcal{B}}_{0} {\mathcal{B}}_{1} + {\mathcal{B}}_{1}^{2}\right)}} \left(-\frac{46}{45}+\frac{5\,\mathcal{B}_0}{18\,\mathcal{B}_1}+\frac{5\,\mathcal{B}_1}{18\,\mathcal{B}_0} \right)+\dots \\
\dots + \frac{56\,\pi\,\mathcal{C}^2\,\mathcal{K}}{5}\,\left(\frac{s_0\,\sin(\mathcal{B}_0\,w)}{\mathcal{B}_0^2}+\frac{s_1\,\sin(\mathcal{B}_1\,w)}{\mathcal{B}_1^2} \right) + \dots\\
\dots + \frac{71\,\mathcal{C}^2}{360}\,\left(\frac{s_0^2\,\cos(2\mathcal{B}_0\,w)}{\mathcal{B}_0^2}+\frac{s_1^2\,\cos(2\mathcal{B}_1\,w)}{\mathcal{B}_1^2} \right)\,.
\end{multline}
As it was expected, the functions $F^{(2)}$ and $B^{(2)}$ are proportional to $\mathcal{C}^2$, so the method of construction of the metric functions is self-consistent. This way we end up with a dust-like solution up to the second order of the perturbation theory.

\subsection{Third and fourth order perturbations.}\label{sec:34}
It is possible to apply the procedure given in the previous subsection in the consecutive orders of the perturbation theory.
First, we fix the values of the powers: 
\begin{equation}\label{Eqn:powers}
\alpha_k=(2\,k)/3, \qquad \beta_k=(2\,k-1)/3, \qquad \phi_k=(2\,k-1)/3\,,
\end{equation}
which appear in the time evolution part of the metric functions. This way we simplify the time dependence of 
the energy-momentum tensor subexpressions to a single power law. Next, we assume that metric functions $B^{(k)}$ 
and $F^{(k)}$ are proportional to $\mathcal{C}^2$ and neglect the terms of the energy-momentum tensor which 
are small in comparison to the leading-order terms. Since $T^{(k)\:\!0}\,{}_i$ and $T^{(k)\:\!i}\,{}_j|_{i\neq j}$
contains only the functions $B^{(k)}$ and $F^{(k)}$ and the constant $\mathcal{C}^2$ is small, we may expect
that $T^{(k)\:\!0}\,{}_i\approx 0$ and $T^{(k)\:\!i}\,{}_j\approx 0$ for $i\neq j$. 

In the remaining elements $T^{(k)\:\!i}\,{}_j|_{i=j}$ one can identify terms depending on two variables, which can
be cancel out when the function $F^{(k)}$ satisfies appropriate differential equation similar to Eqn. \ref{Eqn:F2}.
Then, in the formula for $T^{(k)\:\!i}\,{}_j|_{i=j}$ remain some terms depending on the one variable only.
They can be set to zero, by demanding that the function $B^{(k)}$ satisfies some differential equation similar to
Eqn. \ref{Eqn:B2}.

The differential equations for the metric functions $F^{(k)}$ and $B^{(k)}$ enable one to simplify the formula
for the $k$-order density $\rho^{(k)}$. In effect, $\rho^{(k)}$ depends only on the metric functions $A^{(l)}$, for 
$l\leq k$. This way, specification of the arbitrary metric function $A^{(k)}$ determines the spatial profile of 
the $k$-order density. In our example, we analyze the case for which the spatial distribution of the density 
in each order is the same. This means that:
\begin{equation}\label{Eqn:rhoK} 
\rho^{(k)}=-\mathcal{K}\,t^{(2k-6)/3}\,\left(A^{(1)}(x)+A^{(1)}(y)+A^{(1)}(z) \right)\,,
\end{equation} 
for $k\geq 2$. The time dependence $t^{(2k-6)/3}$ of the $k$-order density is a consequence of the specific values 
of the powers Eq. \ref{Eqn:powers}. 

The right-hand sides of the differential equations for $F^{(k)}$ and $B^{(k)}$ depend on the metric functions
$F^{(l)}$ and $B^{(l)}$ known from the previous orders $l<k$ and the metric functions $A^{(m)}$, for $m\leq k$.
After the arbitrary function $A^{(k)}$ is fixed, one can solve these differential equations and obtain
the resulting $F^{(k)}$ and $B^{(k)}$. It is easy to verify that these functions are proportional to $\mathcal{C}^2$,
so the method is self-consistent.   

We apply the presented method up to fourth order of the perturbation theory. The resulting metric functions
are made of simple trigonometric and monomial functions, but the expressions become quite large and we decided not to display them.

\subsection{Exact solution}\label{sec:exact_solution}
By the method presented in the previous subsections we construct a dust-like solution to the fourth-order 
perturbation theory. Now, we would like to use the solution we obtained so far as a guess of the metric functions
$c_{i\:\!j}(\lambda,t,x,y,z)$ of some exact solution to the Einstein equations. More precisely, we assume that
 the space-time metric of the model universe is a fourth-order polynomial in the parameter $\lambda$:
\begin{equation}\label{Eqn:fullmetric}
 g_{i\:\!j}=\sum\limits_{k=0}^{4} \lambda^k\,g^{(k)}_{i\:\!j}\,,
\end{equation}
where the metric functions in each order $g^{(k)}$ were constructed by the method described in the subsections 
\ref{sec:lin}-\ref{sec:34}. In the next section we verify that the energy-momentum tensor $T_{\mu\:\!\nu}=G_{\mu\:\!\nu}/(8\pi)$
refers to some dust-like matter distribution and we analyze in details the model properties.

\section{Model properties}
\subsection{Density distribution}
Let us begin with the analysis of the distribution of the model density $\rho=-T^{0}\,{}_0$. In the model construction
method presented in the previous section, in each order $k\geq 1$ one can specify the shape of the $k$-order
density distribution $\rho^{(k)}$ by means of arbitrary function $A^{(k)}$. We fix the $A^{(1)}$ function as in the
definition Eq. \ref{Eqn:A1}. The functions $A^{(k)}$ for $1<k\leq 4$ has been specified so that the density $\rho^{(k)}$
has the same spatial distribution as $\rho^{(1)}$. The formula for $\rho^{(k)}$ is given by the Eqn. \ref{Eqn:rhoK}. 
The density in each order $\rho^{(k)}$ has a different power in a time dependence. To some extent, we can control the growth rate of the inhomogeneities by manipulating the contribution of each order $\rho^{(k)}$ to the total density. In our example, the parameter $\mathcal{K}$ describes the contribution of the high-order densities
$\rho^{(k)}|_{1<k\leq 4}$ in comparison to the linear order contribution $\rho^{(1)}$, given by the Eq. \ref{Eqn:rho1}.

From now on, we fix the parameter $\lambda=4.42\times 10^{-4}$. For this value of $\lambda$ the maximal density
at the EdS universe age $\EdStime$ up to the first order is $\Omega^{(0)}+\Omega^{(1)}=1.2$, where $\Omega^{(k)}=\rho^{(k)}/\rho_{cr}$.
For $\mathcal{K}=0$ we expect that the first-order density gives a dominant contribution to the total density. Indeed, the total density expressed in the units of the critical density $\Omega=\rho/\rho_{cr}$
evaluated at the maximum of the overdensity region $\vec{x}_{O}=(12.5,12.5,12.5)$ and at the $\EdStime$ is $\Omega=1.1999$.
The shape of the isodensity surfaces remains practically unchanged in comparison to the isodensity surfaces of the
perturbation theory up to the first order. Therefore, density of the full model $\rho$ is approximated very well
by the density of the perturbation theory up to the first-order perturbations. Also, one can think about Figure \ref{fig:isodensity} as it describes the density distribution in space of the density of the full model $\rho$. We examined also two specific values $\mathcal{K}=0.1$ and $\mathcal{K}=-0.1$, for which the density of the full model is very close to the density of the fourth-order perturbation theory.

\begin{figure}[h]
	\centering
   \includegraphics[width=0.3\textwidth]{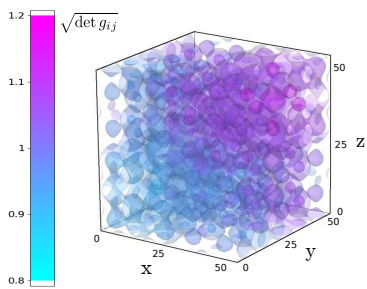}
	\caption{\label{fig:vokumeelement} \scriptsize{The isosurfaces of the geometrical factor $\sqrt{\mathrm{det}g_{i\:\!j}}$ appearning in the volume element $\sqrt{\mathrm{det}g_{i\:\!j}}\,\ud^3x$. The domain is the elementary cell.}}
\end{figure}

The density has a spatial distribution, which forms an infinite, cubic lattice. For a scales much larger than
the elementary cell the model becomes homogeneous and isotropic in common sense. It is then reasonable to approximate
our inhomogeneous universe by the FLRW solution with some average density distribution $\langle\rho\rangle$
if only scales much larger than the elementary cell are considered. We will use a natural definition of the average
of some physical quantity $f$ over the elementary cell $\mathcal{D}$, at the hypersurface of the constant time $t$:
\begin{equation}
\langle f \rangle_{\mathcal{D}}(t)=\frac{\int_{\mathcal{D}}\ud^3 x\,f(t,\vec{x})\,\sqrt{\mathrm{det}g_{i\:\!j}}}{\int_{\mathcal{D}}\ud^3 x\,\sqrt{\mathrm{det}g_{i\:\!j}}}\,.
\end{equation} 
It should be pointed out that the volume element is not trivial and depend on the position in space. The isosurfaces of the square root of the spatial part of the metric determinant are shown in Fig. \ref{fig:vokumeelement}. 
Comparison of this figure with \mbox{Fig. \ref{fig:isodensity}} shows that underdensity regions have larger values
of the volume element than overdensities. 

\begin{figure}[h]
	\centering
   \includegraphics[width=0.45\textwidth]{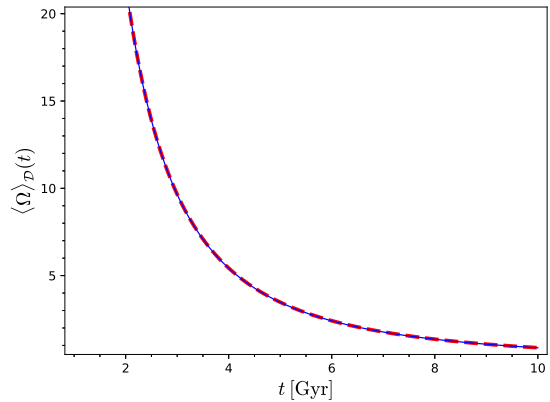}
	\caption{\label{fig:average_rho} \scriptsize{\emph{Blue} - the model average density expressed in the critical
	 density units as a function of time. \emph{Red, dashed curve} correspond to the time dependence of the density
	  of the Einstein-de Sitter model. }}
\end{figure}

In Figure \ref{fig:average_rho} we present by the solid blue curve the model average density over the elementary cell
$\langle \Omega\rangle_{\mathcal{D}}(t)$ as a function of time, calculated for the case $\mathcal{K}=0$.
For two other values of $\mathcal{K}=0.1$ and $\mathcal{K}=-0.1$ we obtain the same result.
 For comparison, by a red, dashed curve we plot on the same figure the density of the Einstein-\mbox{de Sitter} 
 model, which was used as a background space-time $g^{(0)}$ for the model construction.
It is evident that both curves overlap, so the density of the EdS model is indeed an averaged density of the 
full inhomogeneous model considered here.

\begin{figure}[h]
	\centering
   \includegraphics[width=0.45\textwidth]{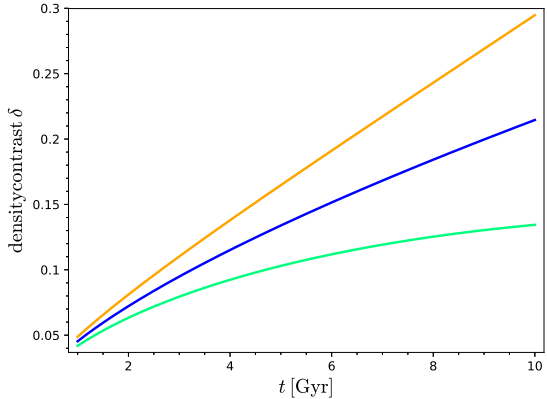}
	\caption{\label{fig:densitycontrast} \scriptsize{The dependence of the density contrast in time. \emph{Blue} 
	curve correspond to $\mathcal{K}=0$, \emph{orange} reffers to $\mathcal{K}=0.1$ and \emph{light green} to
	$\mathcal{K}=-0.1 $.}}
\end{figure} 
The notion of the average density enables one to define the density contrast as:
\begin{equation}
\delta=\max\limits_{x^\mu\in\mathcal{D}}\frac{|\rho-\langle \rho\rangle_{\mathcal{D}}|}{\langle \rho\rangle_{\mathcal{D}}}\,.
\end{equation}
In Figure \ref{fig:densitycontrast} we present the density contrast as a function of time for three values of $\mathcal{K}$. 
For $\mathcal{K}=0$ density contrast grows with time exactly as the growing mode of the first-order perturbation
theory, where $\delta\propto t^{2/3}$. For the parameter $\mathcal{K}=0.1$ density contrast grows faster than $t^{2/3}$,
 while for the value $\mathcal{K}=-0.1$ it grows slower. It is interesting to notice, that the density contrast
of the exact solution to the Einstein equations could differ from the prediction of the first-order perturbation
theory. Important difference appears when second and higher-order terms contribute significantly to the total density.
 
\subsection{Is the energy-momentum tensor dust-like?}
We developed our model within the framework of the perturbation theory up to the fourth order. Then, we treated
the resulting fourth-order polynomial in $\lambda$ parameter (\mbox{Eqn. \ref{Eqn:fullmetric}}) as a metric of the full model, for which the Einstein
equations are satisfied exactly. In effect, we have to deal with the contributions to the energy-momentum
tensor coming from fifth and higher orders, which we do not control. We have to check whether the resulting energy-momentum tensor of the full model $T_{\mu\:\!\nu}=G_{\mu\:\!\nu}/(8\pi)$ resembles the properties of
the energy-momentum tensor from the fourth-order perturbation theory. 

In the previous subsection, we have checked that the density of the full model practically does not change in
comparison with the EdS model density perturbed up to the fourth order. Now, we analyze the values of the other
elements of the energy-momentum tensor. Because of the symmetry condition imposed on the metric, 
there are four types of the $T^{\mu}{}_{\nu}$ components. 
\begin{figure}[h]
	\centering
   \includegraphics[width=0.45\textwidth]{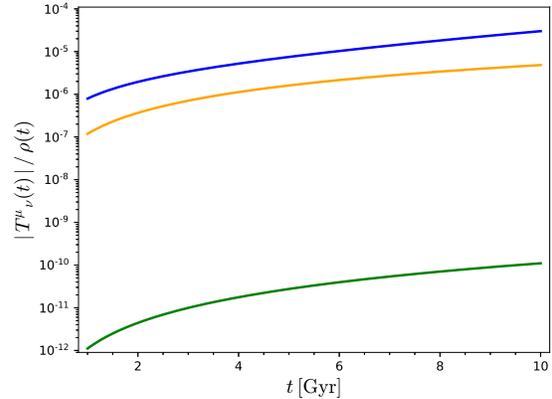}
	\caption{\label{fig:energymomentum} \scriptsize{The elements of the energy-momentum tensor relative to the 
	energy density, evaluated at the center of the overdensity region. \emph{Blue curve} represents pressure-like terms
	$T^{i}{}_{j}|_{i=j}$, \emph{the orange one} corresponds to the elements $T^0{}_i$, while \emph{green curve}
	reffers to $T^{i}{}_{j}|_{i\neq j}$.  }}
\end{figure} 
In Figure \ref{fig:energymomentum} we present the absolute value of the remaining three types of the energy-momentum
elements relative to the density. The energy-momentum tensor elements plotted on this figure are evaluated
at the position of the maximal density $\vec{x}_{O}=(12.5,12.5,12.5)$ and shown as functions of time. The blue curve
corresponds to the elements $T^{i}{}_{j}|_{i=j}$ representing the pressure measured by the observer which is in rest
in the comoving reference frame $(t,x,y,z)$. To verify whether the values of these elements are small enough we
have to consider some interpretation of the source of this pressure.

\begin{figure}[h]
	\centering
   \includegraphics[width=0.3\textwidth]{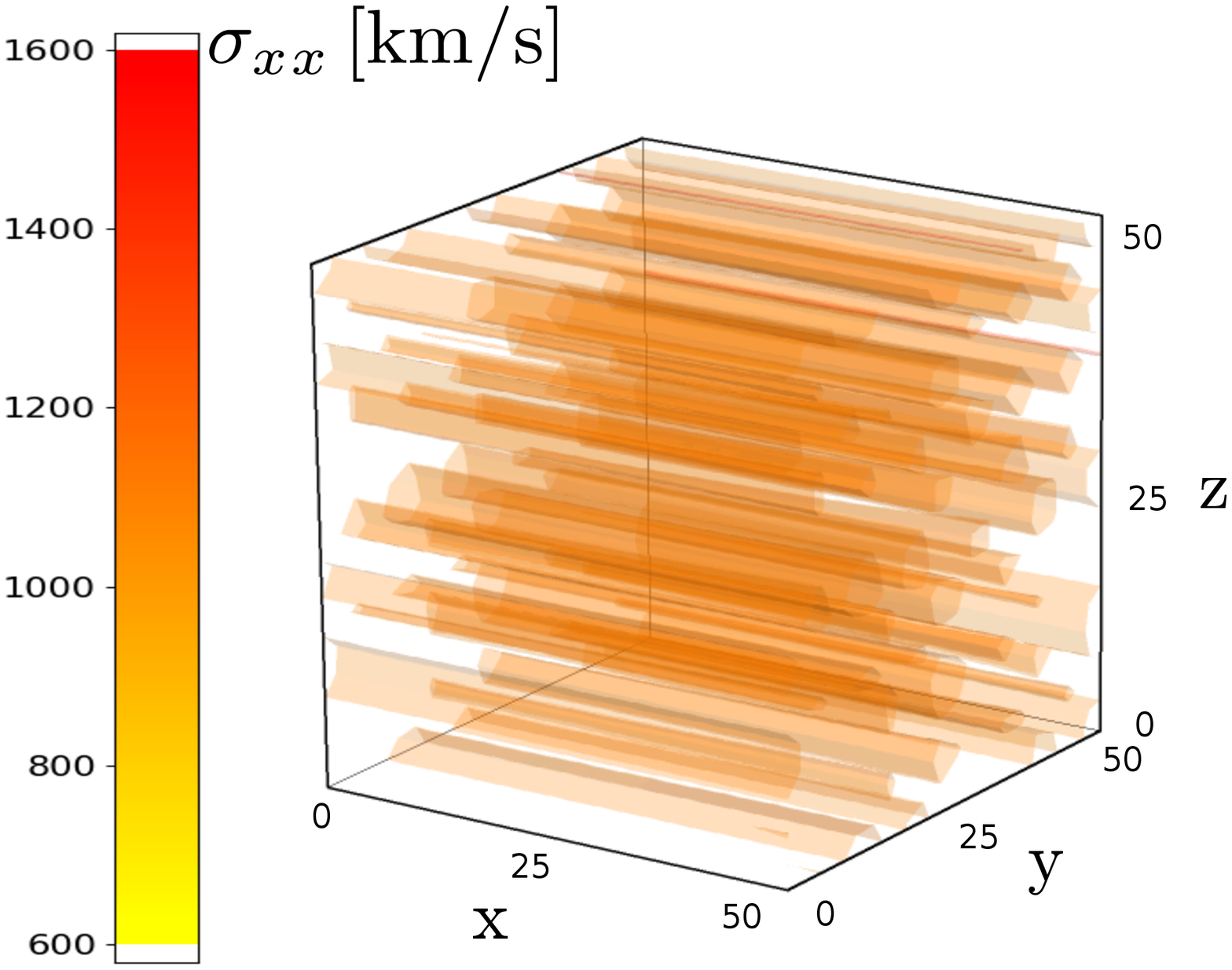}
   \includegraphics[width=0.35\textwidth]{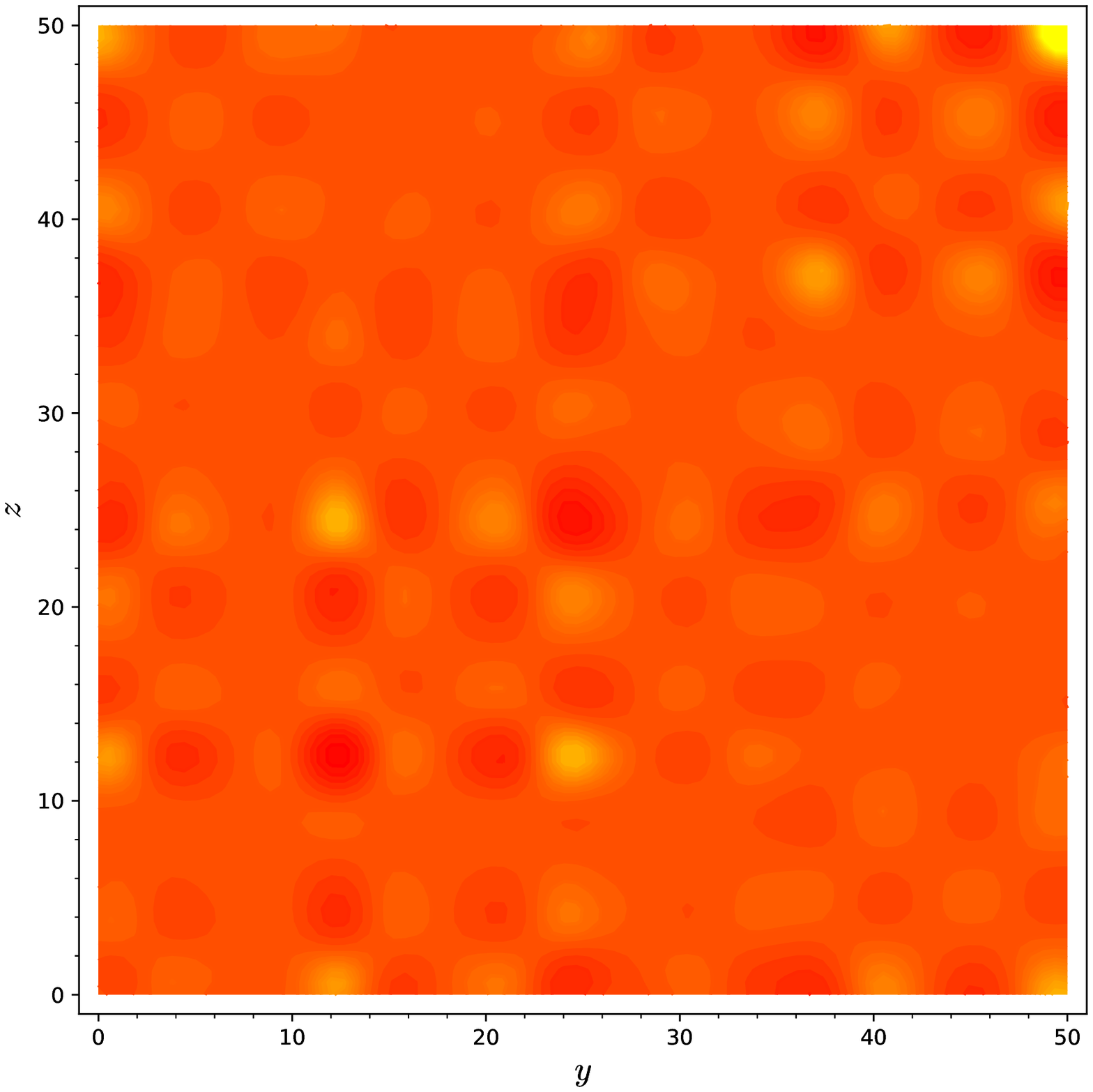}
	\caption{\label{fig:velocitydispersion} \scriptsize{The velocity dispersion element $\sigma_{x\:\!x}$. Top panel:
   isosurfaces of constant $\sigma_{x\:\!x}$ within the elementary cell. Bottom: The cross-section of the $\sigma_{x\:\!x}$
   profile. The colors encode the scale of the velocity dispersion in the same way as in the top panel.   }}
\end{figure}

If the model energy-momentum tensor describes galaxies which are members of rich galaxy clusters and have its
proper motions, then the energy-momentum tensor could be interpreted in the framework of the Jeans theory of a collisionless system of particles. Within this model framework, the stress-energy tensor (the spatial part of
the energy-momentum tensor) is directly related to the velocity dispersion tensor of these particles $\sigma_{i\:\!j}$:
\begin{equation}\label{Eqn:jeans}
T^i{}_j=\rho\,\sigma^2_{i\:\!j}\,.
\end{equation}
Since the density is positive, the elements of $T^i{}_j$ should be positive also. Unfortunately, the resulting
pressure $T^{i}{}_{j}|_{i=j}$ is negative in some regions. However, this problem can be solved in the following way.
We increase a bit the pressure by adding the small positive term $\mathcal{P}\,t^{-2/3}$ to the right-hand side
of the second-order formula Eqn. \ref{Eqn:T2xx}, where $\mathcal{P}=1.006\times10^{-4}$. Then, after recalculation
of the metric functions, one can check that in the case $\mathcal{K}=0$ the order of magnitude of the energy-momentum
tensor elements remain unchanged in comparison to the case $\mathcal{P}=0$, but the pressure $T^{i}{}_{j}|_{i=j}$
is always positive within the elementary cell. For other values of $\mathcal{K}$ the corresponding $\mathcal{P}$ 
should be different. In Figure \ref{fig:velocitydispersion} we present one of the elements of the velocity dispersion
tensor $\sigma_{x\:\!x}$, which is related to the model energy-momentum tensor element $T^{x}{}_{x}$ by the formula
Eqn. \ref{Eqn:jeans}, for the case $\mathcal{K}=0$ and at the time $\EdStime$.
This figure shows that the values of the pressure of order $10^{-6}$ in the geometrized units correspond to the velocity
dispersion of order of $1000\,\mathrm{km/s}$. This value of the velocity dispersion is consistent with observations
of galaxy clusters. The nondiagonal elements $T^{i}{}_{j}|_{i\neq j}$ are very close to zero. The order of magnitude
of these elements is $10^{-10}$. This suggests that the distribution of the velocities should be isotropic. As can be seen
in Fig. \ref{fig:velocitydispersion}, this is not the case of the resulting $\sigma_{i\:\!j}$. However,
the order of magnitude of the spatial part of the resulting energy-momentum tensor $T^{i}{}_{j}$ is consistent
with the values of the velocity dispersion found in the galaxy clusters.

The energy flux terms $T^{0}{}_i$ are one order of magnitude lower than the pressure-like terms $T^{i}{}_{j}|_{i=j}$,
therefore we conclude, that the resulting energy-momentum tensor of the full inhomogeneous solution considered here
is really dust-like.
 
\subsection{Curvature of space.}
The EdS model is the background space-time for the perturbative construction scheme presented in Section \ref{sec:construction}. The EdS universe is spatially flat. In this subsection, we will analyze the behavior of the 
spatial curvature of the full model.

In the Einstein--de Sitter model understood as a special case of the Friedmann--Lema\^itre cosmological model, additionally, there vanish the curvature scalar of hypersurfaces orthogonal to the fluid flow $\mathcal{R}$ and the isotropic pressure $p$. By the Stewart--Walker lemma \cite{1974RSPSA.341...49S}, in the perturbed Einstein--de Sitter model at the first order, these two scalar fields are gauge-invariant. In contrast, the perturbation of the matter energy density $\rho$ is not gauge-invariant but one can consider its spatial gradient $X{_\nu}=P{_\nu}{^\alpha}\nabla{_\alpha}\rho$ as a~suitable perturbative quantity. Geometric, kinematic and dynamic gauge-invariant quantities which characterize properties of the perturbed space-time, energy-momentum field and its flow are mutually related by the Ellis--Bruni equations \cite{1989PhRvD..40.1804E}. In special cases, when this set of equations become closed, it provides analytic solution for the behavior of inhomogeneities. When we restrict our considerations only to perturbations of the scalar type then the fluid flow is necessarily irrotational and the magnetic part of the Weyl tensor vanishes \cite{1989PhRvD..39.2882G}. If we further assume that the perturbed flow is geodesic and the fluid is nonconductive and inviscid then we arrive at the following set of equations for perturbative quantities
\begin{gather}
U{^\alpha}\nabla{_\alpha}\mathcal{R}+\frac{2}{3}\theta\mathcal{R}=0,\\
U{^\alpha}\nabla{_\alpha}(8\pi X{_\nu})+\frac{11}{6}\theta8\pi X{_\nu}-\frac{1}{4}\theta P{_\nu}{^\alpha}\nabla{_\alpha}\mathcal{R}=0,\\
p=0,
\end{gather}
where $P{_\mu}{_\nu}=U{_\mu}U{_\nu}+g{_\mu}{_\nu}$ is the projection tensor and $\theta$ is the expansion scalar in the background. It appears that the imposed assumptions do not allow for nonzero pressure perturbations. Furthermore, they confine only the temporal evolution of perturbations leaving their spatial variability free. Curvature perturbations decrease with the expansion of the model but solving for density perturbations reveals two separate modes, one decaying and one growing. These modes differ in their physical nature, since the growing mode is govern entirely by the curvature perturbations of hypersurfaces. In particular, imposing zero curvature perturbations on hypersurfaces eliminates the growing mode of density perturbations.

\begin{figure}[h]
	\centering
   \includegraphics[width=0.3\textwidth]{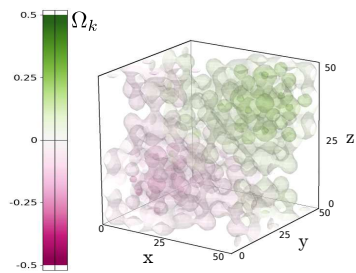}
	\caption{\label{fig:omegaK} \scriptsize{The isosurfaces of constant $\Omega_k$ parameter evaluated at the
    age of the EdS universe $\EdStime$. The domain is the elementary cell.}}
\end{figure}

After these general considerations, let us go back to the specific exact solution presented in Sec \ref{sec:exact_solution}. The scalar curvature of the hypersurface of a constant time $\mathcal{R}$ is conventionally related to the quantity
\mbox{$\Omega_k=-\mathcal{R}/(6\,H(t)^2)$}, where $H(t)$ is the Hubble parameter. We used the Hubble parameter $H(t)$
of the background EdS universe. In Figure \ref{fig:omegaK} we show a dependence of the $\Omega_k$ on the position 
within the elementary cell, at the time $\EdStime$. The overdense regions have negative values of $\Omega_k$, so 
the scalar curvature is positive in these regions. Within the underdense regions the situation is opposite. 
The $\Omega_k$ parameter is positive and the scalar curvature $\mathcal{R}$ is negative there.

\begin{figure}[h]
	\centering
   \includegraphics[width=0.45\textwidth]{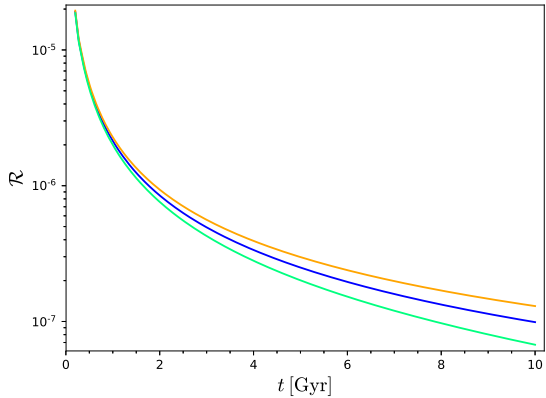}
	\caption{\label{fig:R3} \scriptsize{The scalar curvature of the hypersurface of a constant time $\mathcal{R}$
    as a function of time. We plot three cases with a different $\mathcal{K}$ parameter. \emph{Blue} curve for 
    $\mathcal{K}=0$, \emph{orange} curve corresponding to $\mathcal{K}=0.1$ and \emph{light-green} for the value
	$\mathcal{K}=-0.1 $.  }}
\end{figure}

Let us analyze the dependence of the scalar curvature $\mathcal{R}$ on time. In Figure \ref{fig:R3} we plot 
$\mathcal{R}$ evaluated at the position $\vec{x}_{O}=(12.5,12.5,12.5)$ of the maximal density.
It is seen that the scalar curvature decreases with time and space tends to flatten during the time evolution.
The orange curve corresponds to the model with the value $\mathcal{K}=0.1$, while the light-green curve refers
to $\mathcal{K}=-0.1$. We note that for the case $\mathcal{K}=0.1$, for which the growth rate of the inhomogeneities
is higher than for the case $\mathcal{K}=-0.1$, the scalar curvature $\mathcal{R}$ decreases more slowly than
for the case with $\mathcal{K}=-0.1$. This shows, that the growth rate of the inhomogeneities is related to the scalar
curvature $\mathcal{R}$ behavior. 

\begin{figure}[h]
	\centering
   \includegraphics[width=0.45\textwidth]{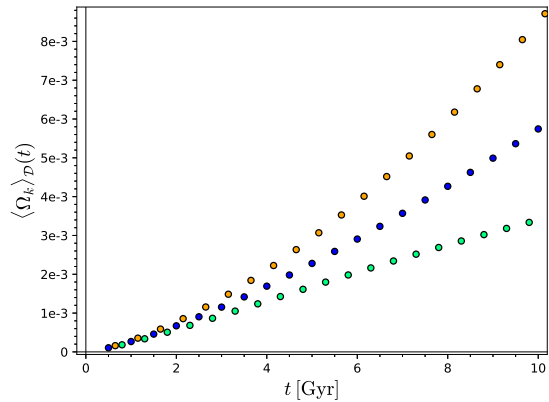}
	\caption{\label{fig:averageOmegaK} \scriptsize{Average $\Omega_k$ parameter as a function of time. As on the previous
    plots, the \emph{blue} points correspond to $\mathcal{K}=0$, the \emph{orange} ones to $\mathcal{K}=0.1$ and 
    \emph{light-green} points reffer to $\mathcal{K}=-0.1 $.  }}
\end{figure}

Finally, we calculate the average over the elementary cell of the parameter $\Omega_k$. The results for three different
values of $\mathcal{K}$ and for different instants of time are plotted on Fig. \ref{fig:averageOmegaK}.   
In the paper \cite{2018PhRvD..97j3529B}, the authors based on their silent universe model suggest that the scalar curvature of space
could emerge with time. In our case, we observe that the average $\Omega_k$ parameter grows slightly with time, 
however, values of $\langle\Omega_k\rangle_{\mathcal{D}}$ are very small. The growth rate of $\langle\Omega_k\rangle_{\mathcal{D}}$
depends on the $\mathcal{K}$ parameter, but in each case, the values of the $\langle\Omega_k\rangle_{\mathcal{D}}$ are smaller than
$10^{-2}$. This means that on average the space is almost flat.
  
\subsection{Local measurments of the Hubble constant.}
\begin{figure}[h]
	\centering
   \includegraphics[width=0.45\textwidth]{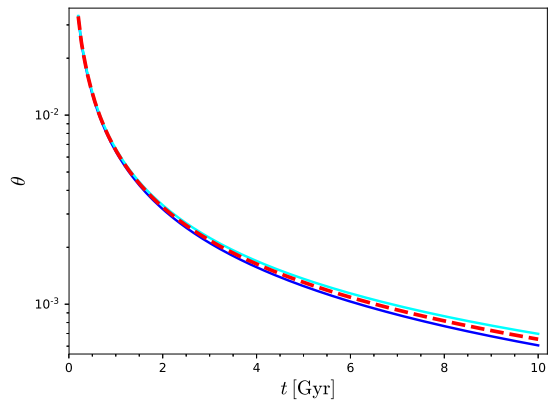}
	\caption{\label{fig:theta} \scriptsize{Expansion scalar $\theta$ as a function of time. The quantity $\theta$ is evaluated
    at the overdensity - \emph{blue curve} or at the underdensity - \emph{light-blue curve}. For comparison, the expanssion
    scalar of the EdS universe is plotted by \emph{red dashed curve}. }}
\end{figure}

At the end of this paragraph let us analyze expansion of our inhomogeneous universe. In Figure \ref{fig:theta} we plot
the expansion scalar $\theta=-K^{i}{}_i$ as a function of time, where $K^{i}{}_j$ is the extrinsic curvature tensor of 
the hypersurface of a constant time $t$. The blue curve represents the expansion scalar at the position of the maximum density 
$\vec{x}_{O}=(12.5,12.5,12.5)$, while the light-blue curve correspond to $\theta$ evaluated at the position of the minimum 
density $\vec{x}_{U}=(37.5,37.5,37.5)$. The red dashed curve shows the expansion scalar of the EdS universe.
On the basis of this figure we deduce that underdense regions expand faster than overdense regions, although on average
the model universe expand exactly as the Einstein-de Sitter homegeneous case. Therefore, local measurments of the Hubble 
constant could differ from the EdS Hubble parameter $H(t)$ evaluated at the universe age $\EdStime$, while the measurements
of the Hubble constant on basis of some obervables related to scales much larger than the elementary cell should be consistent
with the EdS prediction. 

To simulate how observer living in our inhomogeneous universe would perform local measurement of the Hubble constant,
we made the following numerical experiment. For a given observer position $\vec{x}_0=(x_0,y_0,z_0)$ we generate ten random directions
$(\theta,\phi)$ with probability distribution uniform on the unit sphere. For each direction we generate randomly ten points
belonging to the line $\gamma(l)=(\EdStime,x_0+l\,\sin\theta\cos\phi, y_0+l\,\sin\theta\sin\phi, z_0+l\,\cos\theta)$. 
This way we simulate one hundred sources distributed randomly in the close neighborhood of the observer. To generate the Hubble
diagram we have to calculate for each source the physical distance to the observer $d$ and its time derivative $\dot{d}$.
The physical distance we obtain by numerical integration: 
\begin{equation}
d(\widetilde{l})=\int\limits_{0}^{\widetilde{l}}\sqrt{\gamma'(l)^i\,\gamma'(l)^j\,g_{i\:\! j}}\,\mathrm{d}l\,.
\end{equation}
Since the metric elements depend explicitly on time, to get the respective $\dot{d}$ we need to take the time derivative of the
integration kernel and perform numerical integration again. 

\begin{figure}[h]
	\centering
   \includegraphics[width=0.45\textwidth]{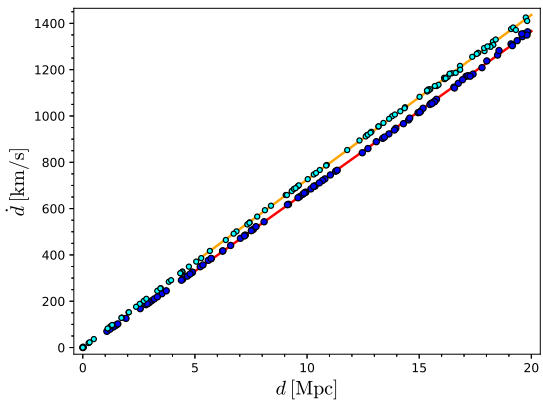}
	\caption{\label{fig:Hubble} \scriptsize{The Hubble diagram generated for the observer located at the overdensity - \emph{blue points}
    and for the observer at the underdensity - \emph{light-blue points}. The \emph{red} and \emph{orange} lines are the respective
    linear fits to the generated points. }}
\end{figure} 
The resulting Hubble diagram generated for two different observer's positions is plotted in Figure \ref{fig:Hubble}.
The blue points correspond to the observer located at the maximum density $\vec{x}_{O}$, whereas the light-blue points are generated
for the observer located at the minimum density $\vec{x}_{U}$. For the points in the range of distances $d\in(5,20)\,\mathrm{Mpc}$
we perform the linear fit to get the resulting local Hubble constant. For the observer located at the overdensity $\vec{x}_{O}$ 
we obtain $H_0=69.12\,\mathrm{km/s/Mpc}$, while the observer located at the underdensity measures $H_0=71.11\,\mathrm{km/s/Mpc}$. It is clear that in both cases the resulting value differs
significantly from the EdS Hubble constant $H_0=70\,\mathrm{km/s/Mpc}$, which was fixed at the beggining of the perturbative approach described in Sec. \ref{sec:construction}. The difference we obtained here is slightly lower than the current observational difference between the Hubble constant estimation from CMB $H_0=67.37 \pm 0.54\,\mathrm{km/s/Mpc}$ \cite{2018arXiv180706209P} and from the local measurements $H_0=74.03 \pm 1.42 \,\mathrm{km/s/Mpc}$ \cite{2019ApJ...876...85R}. However, the order of magnitude of the difference we get here is comparable to the current observational difference. It is then reasonable to expect that inhomogeneities could play an important role concerning local Hubble constant measurements. 

In our previous paper \cite{2019PhRvD..99h3521S} we haven't noticed a dependence of the local measurement of $H_0$ on the position
within the elementary cell. However, in the previous model we considered smaller scale of the inhomogeneities, of the order of $3\,\mathrm{Mpc}$. Currently, the scale of the inhomogeneous region is close to $25\,\mathrm{Mpc}$ with the additional substructures of the scale around $3\,\mathrm{Mpc}$. It is quite clear that the local measurements of the Hubble constant should
depend on the scale of the inhomogeneities under consideration.

\section{Conlusions.}
In the current paper, we constructed an example of the dust-like exact solution to the Einstein equations representing 
an inhomogeneous cosmological model with growing amplitude of the inhomogeneities. By the term dust-like we mean that an observer
which is in rest in the comoving reference frame measures nonzero energy-momentum tensor elements $T^{\mu}{}_{\nu}$ other than the energy density 
$-T^{0}{}_{0}$, but which are negligible in comparison to $-T^{0}{}_{0}$. In the interpretation of the Jeans theory of the collisionless
system of particles as a source of the pressure-like terms of the energy-momentum tensor, we checked that the values of the resulting 
energy-momentum tensor elements correspond to the particles velocity dispersion around $1000\,\mathrm{km/s}$,
which is reasonable value for the velocity dispersion of the galaxy cluster members.

The model construction method is based on the perturbation theory around the Einstein-de Sitter background. By using the 
additional simplifying symmetry condition and identifying leading terms of the energy-momentum tensor elements we are able
to obtain the solution to the perturbation theory up to the fourth-order perturbations. Then, we consider the resulting
fourth-order solution as a guess of the metric of some exact solution to the Einstein equations. We checked that this solution
remains dust-like and analyze its basic properties.

Many of the current discussions concerning the possible influence of the inhomogeneities on the cosmological observations focus
on the problem of backreaction. In these approaches, one asks whether the existence of the inhomogeneities affects properties
of the average space-time. Recently, most of the researchers conclude that the effect of backreaction is possible but it is rather
negligible. In the presented model there is no backreaction effect at all. If we consider Eqn. \ref{Eqn:fullmetric}
as a definition of the Green-Wald family of metrics $g_{\mu\:\!\nu}(\lambda)$ \cite{2011PhRvD..83h4020G}, then the effective energy-momentum tensor $t_{\mu\:\!\nu}$
is zero since our model is based on the ordinary perturbation theory. Moreover, the average over the whole space of the density
$\langle\rho\rangle_{\mathcal{D}}(t)$, the average curvature parameter $\langle\Omega_k\rangle_{\mathcal{D}}$ and the average
expansion $\langle\theta\rangle_{\mathcal{D}}$ overlap with the respective quantities of the background EdS model. At the same time, local physical quantities could differ significantly from the EdS background. We note nonnegligible differences considering the local 
volume element $\sqrt{\mathrm{det}g_{i\:\!j}}\,d^3 x$, local curvature of space and local expansion parameter $\theta$. 
In effect, we show that some local measurements could differ from the expectations of the EdS background model. As the example,
we verify that the local measurements of the Hubble constant could differ from the EdS value. Such an effect could possibly
explain the current Hubble tension problem.

We want to stress, that our model is not a complete description of the real Universe. It is rather a step toward understanding
the role of the large-scale inhomogeneities in the interpretation of the cosmological observables. The present paper is a large
improvement of our previous work \cite{2019PhRvD..99h3521S,2017PhRvD..95f3517S}, but many issues could be done better in the future.
First of all, there is a much more challenging task of considering perturbations around the nonflat background with nonzero
cosmological constant. One can also look for more complicated density distributions beyond our symmetry restrictions. We point also,
that in the present model the density, the scalar curvature of space and the expansion parameter are decreasing functions of time
in all spatial positions. Therefore, the presented model could be interpreted as a description for a large scale inhomogeneous regions 
behaviour but does not provide the framework for the formation of the individual structures for which we expect some kind of collapse 
and very high values of the density contrast. Nevertheless, the model gives an explicit example of an important influence of 
the inhomogeneities on the local observations of the Hubble constant, while at the same time the observables related to the scales 
much larger than the elementary cell overlap with the prediction of the background homogeneous model.

\bibliography{SikoraGlod2019}
\bibliographystyle{unsrt}

\end{document}